\documentclass[prd,twocolumn,showpacs,preprintnumbers,nofootinbib]{revtex4}
\usepackage{amsfonts,amsmath,amssymb}
\usepackage{graphicx}
\usepackage{color}
\usepackage{natbib}
\usepackage{siunitx}
\usepackage{threeparttable}

\usepackage[plainpages=false, colorlinks=true, anchorcolor=blue, linkcolor=blue, citecolor=blue, bookmarks=false]{hyperref}
\newcommand{\rthis}[1]{\textcolor{black}{#1}}

\usepackage{natbib}

\begin{document}
%\markboth{Authors' Names}

%%%%%%%% Journals %%%%%%%%%%%%%%

%%%%%%%%%%%%%%%%%% TITLE %%%%%%%%%%%%%%%%%%%%%%%%%%%%%%%%%%%% 
\title{A meta-analysis of  neutron lifetime measurements}
\author{Ashwani \surname{Rajan}$^1$}  \altaffiliation{E-mail: ashwani.rajan135@gmail.com}
\author{Shantanu  \surname{Desai}$^2$} \altaffiliation{E-mail: shntn05@gmail.com}

\affiliation{$^{1}$Department of Physics, Indian Institute of Technology, Guwahati, Assam-781039, India}
\affiliation{$^{2}$Department of Physics, Indian Institute of Technology, Hyderabad, Telangana-502285, India}

\begin{abstract}
We calculated the  median as well as weighted mean central estimates for the neutron lifetime, from  a subset of  measurements compiled in the  2019 update of the  Particle Data Group (PDG). We then reconstruct the error distributions for the  residuals using three different central estimates and then check for  consistency with a Gaussian distribution.  We find that although the error distributions using the weighted mean as well as median estimate are consistent with a Gaussian distribution, the Student's $t$ and Cauchy distribution provide a better fit. This median statistic estimate of the neutron lifetime from these measurements is given by $881.5 \pm 0.47$ seconds. \rthis{This can be used as an alternate  estimate of the neutron lifetime.} %However, using simulations we showed that with 19 data points, the $p$-values we obtain are consistent for data with Gaussian error distributions.
We also note that the  discrepancy between beam and bottle-based measurements using median  statistics of the neutron lifetime persists with a significance between 4-8$\sigma$, depending on which combination of measurements is used.
\end{abstract}

\maketitle

%%%%%%%%%%%%%%%%%%%%%%%%%%%%%%%%%%%%%%%%%%%%%%%%%%%%%%%%%%%
%%%%%%%%%%%%%%%%%%%%%%%%%%%%%%%%%%%%%%%%%%%%%%%%%%%%%%%%%%%
%%%%%%%%%%%%%%%%%%%%%%%%%%%%%%%%%%%%%%%%%%%%%%%%%%%%%%%%%%%
%\noindent {\it Introduction.---}
\section{Introduction}
The precise measurement and theoretical estimate of the neutron lifetime is of paramount importance for both particle physics and astrophysics~\citep{neutronlifetime11,neutronlifetime14}. The current weighted average of seven   neutron lifetime measurements, reported in the 2019 version of  the Particle Data Group~\citep{PDG} (PDG, hereafter)\footnote{At the time of writing, the 2019 PDG update on neutron lifetime measurements is only available online at \url{http://pdg.lbl.gov/2019/listings/rpp2019-list-n.pdf}. The published version~\cite{PDG} contains listings from 2018.} using seven best measurements is   $879.4 \pm 0.6$ seconds. At face value, the weighted mean error from these measurements is equal  to 0.4 seconds. Therefore, the reduced  $\chi^2$ value for a constant neutron lifetime  is equal to 14.6  for six degrees of  freedom, corresponding to a $p$-value of 0.023~\citep{NR}. If we define  the significance as the number of standard deviations a Gaussian variable would fluctuate in one direction corresponding to this $p$ value, then the observed $p$-value  corresponds to a  $2\sigma$~\citep{Ganguly} discrepancy for a constant value of the neutron lifetime. Therefore, the PDG has scaled the weighted mean error by a  scale factor  equal to $\sqrt{\chi^2/\nu}$, where $\nu$ is the total degrees of freedom. With this multiplicative scale factor of 1.6,   the total error is now equal to the reported value of 0.6 seconds.  Therefore, the subset of neutron lifetime measurements vetted by the PDG are inconsistent with a constant value at 2$\sigma$ significance. 

\rthis{The theoretical neutron lifetime is a function of the axial vector to vector coupling ratio as well as the CKM matrix element $V_{ub}$~\cite{Fornal,Marciano}. The most recent theoretical estimate of the neutron lifetime is between 875.3 and 891.2 seconds,  within $3\sigma$~\citep{Fornal}. Theoretical uncertainties in the neutron lifetime calculation, and expected improvements  in the near future have been recently reviewed in Ref.~\cite{Marciano}.}

Neutron lifetime measurement techniques can be broadly classified into two types: `bottle' and `beam' based measurements.
%\footnote{We note however that sometimes the bottle-based experiments have also been sub-divided into two further sub-categories depending on the type of bottle used, and are usually subdivided into magnetic bottle-based measurements  and non-magnetic bottle-based measurements~\cite{neutronlifetime11}.} 
In the bottle method, ultra-cold neutrons are stored in a  container (which consists of either some bottle or a trap), and the neutron lifetime is measured by fitting the surviving neutrons to a decaying exponential. In the beam method on the other hand, the number of neutrons and protons are produced from $\beta$-decay, and the lifetime is obtained from the neutron decay rate.
More details about these techniques can be found in Refs.~\cite{neutronlifetime11,neutronlifetime14}.

However, there is a long standing discrepancy between these two methods used for neutron lifetime measurements~\citep{Greene}. As of 2018, the current value  from  two beam experiments~\cite{Byrne96,Yue13} included in the 2018 edition of PDG~\footnote{These two measurements are not used for the neutron lifetime estimate by the 2019 PDG edition.}  is equal to   $888 \pm 2.0$ seconds~\citep{Fornal}, and the same from five bottle experiments~\citep{Mampe,Serebrov05,Pichlmaier10,Steyerl2,Arzumanov15} is  equal to $879.6 \pm 0.6$  seconds~\citep{Fornal}. This is a formally a 4$\sigma$ discrepancy, and as pointed out in Fornal and Grienstein.~\cite{Fornal} (F18 hereafter) could either be evidence of uncontrolled systematics or could point to new physics. Another possibility  however not mentioned  in the above works is that the measurements could contain non-Gaussian errors,  and consequently the weighted mean cannot be used as  the central estimate.

The central estimate of the neutron lifetime  mentioned in PDG as well as all other works, which analyze this discrepancy has been obtained from a weighted average of all the measurements. \rthis {The central estimate of a quantity using weighted measurements makes the following  main assumptions~\cite{Gott}: (i) individual data points are statistically independent and contain no systematic effects ; (ii) the errors are Gaussianly distributed. If any of the measurements contain catastrophic outliers or unaccounted systematic effects, then the second assumption is automatically violated. In that case, the weighted mean can produce extremely biased results. On the other hand, median statistics does not incorporate the individual measurements errors, and hence is unaffected by the presence of a few outliers. Secondly, even if the errors are not correctly estimated, as shown using simulations of Zeldovich's thought experiment involving watches~\cite{Bethapudi}, median estimate gives a more robust estimate.  Even if a dataset is drawn from a distribution with infinite variance such as Cauchy distribution, the median is a more robust central estimate~\cite{Gott}. Many additional pitfalls in using the weighted mean as a central estimate, and how using the median value ameliorates these problems can be found in Refs.~\cite{Gott, Bethapudi} and references therein. The only assumption used for median statistic based estimate is that the measurements are independent and free of systematic errors.}

In the last decade,  Ratra and collaborators have shown that the error distributions for a whole slew  of  astrophysical and cosmological measurements  are inconsistent with a Gaussian distribution~\citep{Gott,Chen03,Chen,Ratra03,Ratra15,RatraLMC,Crandall,Bethapudi,Rajan,RatraD2,Ratratheta}. The datasets they explored for this purpose include measurements of $H_0$~\cite{Ratra03}, Lithium-7 measurements~\cite{Ratra15} (see also ~\cite{Zhang}), distance to LMC ~\cite{RatraLMC}, distance to galactic center~\cite{RatraGC}, Deuterium abundance~\cite{RatraD2}, etc. 
For each of these datasets, they have fit the data to a variety of probability distributions.  From all  these studies, they inferred that the error distribution is non-Gaussian. Consequently, they have argued that median statistics should be used for the central estimates of these parameters instead of the weighted mean~\citep{Gott,Bethapudi}. 
To the best of our knowledge, no one has investigated the Gaussianity of the neutron lifetime measurements (or for that matter any other datasets in PDG). The importance of doing such tests has been stressed in a number of works~\cite{Gott,Crandall,Rajan,Bailey}. 
Due to the non-Gaussanity of the error residuals for the aforementioned astrophysical datasets, median statistics  has   been used to obtain central estimates of some of these quantities such as Hubble Constant~\cite{Gott,Chen,Bethapudi}, Newton's Gravitational Constant~\cite{Bethapudi}, mean matter density~\cite{Chen03}, and other cosmological parameters~\cite{Crandall}. Alternately, one can use the method recently proposed by Cowan, where the uncertainity in the systematic errors has been modeled using probabilistic distributions~\cite{Cowan18}.

Given the importance of the  physics implications of these discrepancies in the neutron lifetime measurements, and to obtain a more robust estimate, which can be easily compared with the theoretical estimate, we revisit the issue of checking for non-Gaussianity of the errors and to obtain a more robust central estimate
from the vetted measurements in PDG. The outline of this manuscript is as follows. The dataset used for our analysis is described in Sect.~\ref{sec:dataset}. Our analysis procedure and results are described in Sect.~\ref{sec:analysis}.  %Tests of validity of $p$-values using mock datasets are discussed in Sect.~\ref{sec:synthetic}. 
We discuss discrepacy between beam and bottle-based measurements in Sect.~\ref{sec:disc}.
We conclude in Sect.~\ref{sec:conclusions}.

\section{Neutron lifetime data}
\label{sec:dataset}
We briefly review the neutron lifetime measurements used for this analysis.  
The 2019 edition of PDG lists a total of 27 measurements from 1972 to present.  From these  measurements, only  seven  have been used by the PDG to obtain the central estimate. Using these seven measurements,  a weighted mean central value of $879.4 \pm 0.6$ s  was estimated, wherein the error has been rescaled by a factor of 1.6. All of these are bottle-based experiments. The corresponding value from the 2018 PDG edition  was  $880.2 \pm 1.0$ s, with five of them been bottle-based and two beam-based.
The remaining measurements were ignored either because the error bars for some of the pre-1980 measurements were large, or if the results from the old measurements were reanalyzed, and lastly because some of the measurements were withdrawn. However, a few measurements have also been culled without any explanation. For our analysis, we also include all  older measurements, except if they were reanalyzed or withdrawn. We also include one   additional measurement~\citep{Leung}, which was not included in either the 2018 or 2019 PDG. In all, we have collected a total of 19 measurements for our analysis, which  are tabulated in Table~\ref{tab:1}. We note that in addition to these direct experimental measurements of neutron lifetime,  there are also cosmological constraints on  the measurements of neutron lifetime~\citep{Salvati}. But we do not include them for our analysis, as these results  are model-dependent, and not direct experimental measurements.

%They also wanted to see if there is evidence for publication bias or ``bandwagon'' effect. 

\section{Analysis}
\label{sec:analysis}

The first step in analyzing the Gaussianity of the error measurements of a dataset is to obtain a central estimate using the available data. For this analysis, we use all the 19 measurements tabulated in Table~\ref{tab:1}. We do not check for Gaussianity  of  the beam and bottle-based measurements separately, as the total number of data points in each category is too small for a robust test. However, once the number of measurements in each category grows, this should also be tested to check for systematics in each category. We note that in P18, a similar analysis was done using 15 deuterium abundance measurements.
Similar to the works by Ratra et al (eg. Ref. ~\citep{RatraD2}, P18 hereafter), we consider two central estimates: weighted mean
and the median. For this analysis, we use all the 19 measurements tabulated in Tab.~\ref{tab:1}.

The median value ($\tau_{med}$) corresponds to the 50\% percentile value, for which half of the data points are below and half above. The standard deviation of the median depends upon the distribution  from where it is sampled from. A number of methods have been proposed in literature to calculate the sample variance of the median~\cite{Woodruff,Maritz,Price}. For this work, to estimate the 68\% confidence interval on the median, we use the methodology   in P18, based on Gott et al~\cite{Gott}, as the estimate is made using only the data and is independent of the sampling distribution.
The weighted mean central value  ($\tau_{wm}$) using the observed neutron lifetime measurements ($\tau_i$) is given by~\cite{Bevington}:
\begin{equation}
\tau_{wm} = \frac{\sum \limits_{i=1}^N \tau_i/\sigma_i^2}{\sum \limits_{i=1}^N 1/\sigma_i^2},
\end{equation}
\noindent where  $\sigma_i$ denotes the total error in each measurement. The total weighted mean error is given by 
\begin{equation}
\sigma_M^2 = \frac{1}{\sum \limits_{i=1}^N 1/\sigma_i^2}. 
\end{equation}

%SD : please fill in these numbers
From the measurements in Table~\ref{tab:1}, the weighted mean estimate is found to be $\tau_{wm} = 879.97 \pm 0.39$ seconds,  and the median estimate is calculated to be $\tau_{med} = 881.5\pm 0.47$ seconds.

\subsection{Error Distributions}
\label{sec:analysisA}
Once we have a central estimate for the neutron lifetime ($\tau_{CE}$) using one of the above three methods, we calculate the residual  error    using~\cite{RatraD2,RatraGC}
\begin{equation}
N_{\sigma_i} =\frac{\tau_i-\tau_{CE}}{\sqrt{\sigma_i^2+\sigma_{CE}^2}}
\label{eq:nsigma}
\end{equation}

\noindent In the above equation, $\sigma_{CE}$ is the error in the central estimate and $\sigma_i$ is the error in the individual measurement. Similar to Refs.~\cite{RatraD2,RatraGC,Ratratheta}, we denote our error distribution for the median ($\tau_{med}$) and the  weighted mean($\tau_{wm}$) calculated from Eq~\ref{eq:nsigma} by $N_{\sigma_i}^{med}$ and  $N_{\sigma_i}^{wm+}$ respectively.  If the central  estimate is determined from the weighted mean, one must also account for correlations and the modified version of the error distribution, which accounts for these correlations is given by~\cite{RatraGC}
\begin{equation}
N_{\sigma_i}^{wm-} =\frac{\tau_i-\tau_{CE}}{\sqrt{\sigma_i^2-\sigma_{CE}^2}}
\end{equation}

Each of these three sets  of  $|N_{\sigma}|$ histograms is then symmetrized around zero.
We now fit the symmetrized histogram of $|N_{\sigma_i}|$ to multiple probability distributions as described in the next section.

\begin{table*}
\begin{tabular}{|l|c|c|c|}
\hline
Reference & Neutron Lifetime (secs) & Type & Comment \\
\hline 
Ezhov 18~\cite{Ezhov18} & $878.3 \pm 1.6   \pm 1.0$  & Bottle & Only in  PDG19 \\ \hline
Serebrov 17~\cite{Serebrov17} & $881.5 \pm  0.7  \pm  0.6$  & Bottle & Only in PDG19 \\ \hline
Pattie 17~\cite{Pattie} &  $877.7 \pm 0.7  + 0.4/-0.2$   & Bottle & Only in  PDG19 \\ \hline
Leung 16~\cite{Leung} & $887 \pm 39$  & Bottle & Neither  PDG18 nor PDG19 \\ \hline
Arzumanov 15 ~\cite{Arzumanov15} & $880.2 \pm 1.2$   & Bottle  &  PDG \\ \hline
Yue 13~\cite{Yue13} &  $887.7 \pm 1.2 \pm 1.9$  & Beam  &  Only in PDG18 \\ \hline
Steyerl 12~\cite{Steyerl2} & $882.5 \pm  1.4 \pm 1.5$   & Bottle  &  PDG \\ \hline
Pichlmaier 10~\cite{Pichlmaier10} & $880.7 \pm  1.3 \pm  1.2$ & Bottle & PDG \\ \hline
Serebrov 05~\cite{Serebrov05} & $878.5 \pm  0.7 \pm  0.3$ & Bottle & PDG \\ \hline
Byrne 96~\cite{Byrne96} & $889.2\pm 3.0 \pm 3.8$   & Beam  &  Only in PDG18 \\ \hline
Mampe 93~\cite{Mampe} & $882.6 \pm 2.7$ & Bottle & PDG \\ \hline
Alfikmenov 90~\cite{Alfimenkov90} & $888.4 \pm 2.9$ & Bottle & PDG (but not used) \\ \hline 
Kossakowski 89~\cite{kossakowski89} & $878 \pm 27 \pm 14$ & Beam & PDG (but not used) \\ \hline  
Paul 89~\cite{Paul89} & $877 \pm 10$ & Bottle & PDG (but not used) \\ \hline
Last 88~\cite{Last88} & $876 \pm 10 \pm 19$ & Beam & PDG (but not used) \\ \hline
Spivak 88~\cite{Spivak88} & $891 \pm 9$ & Beam & PDG (but not used) \\ \hline
Kosvintsev 86~\cite{Kosvintsev86} & $903 \pm 13$ & Bottle & PDG (but not used) \\ \hline
Kosvintsev 80~\cite{Kosvintsev80} & $875 \pm 95$ & Bottle & PDG (but not used) \\ \hline
Christensen 72~\cite{Christensen72} & $918 \pm 14$ & Beam & PDG (but not used) \\ 
\hline
\end{tabular}
\caption{Summary of the 19 measurements used for the analysis. PDG18 refers to the 2018 published version of PDG, and PDG19 refers to the 2019 online update. The last eight are listed in  PDG, but not used to calculate the weighted mean neutron lifetime by both the  PDG editions~\cite{PDG}. The first three measurements are used only the 2019 edition to calculate weighted average. The two beam-based measurements~\cite{Byrne96,Yue13} are only used for the 2018 PDG estimate.}
\label{tab:1}
\end{table*}

\subsection{Fits to probability distributions}
\label{sec:prob}
We fit the symmetrized  histograms for each of the  $|N_{\sigma}|$'s to a Gaussian  distribution as well as  to variants of Gaussian distributions, such as Cauchy, Laplacian, and Student's $t$ distribution, to see which of these  is most compatible with the data. This is  similar in spirit  to recent works by Ratra et al, such as P18 and references therein.  We briefly review this procedure. More details can be found in P18.

The Gaussian distribution we consider has  zero mean and  standard deviation equal to unity
\begin{equation}
P(N) = \frac{1}{\sqrt{2\pi}}\exp(-|N|^2/2)
\label{eq:gauss}
\end{equation}

The second distribution we consider is the Laplacian distribution, which has a  sharp peak and longer tails than a Gaussian distribution and is described by
\begin{equation}
P(N) = \frac{1}{2}\exp(-|N|)
\label{eq:laplace}
\end{equation}

The third distribution, which we will use is the Cauchy or Lorentz distribution. It has longer and thicker tails compared to a  Gaussian distribution. It is described by
\begin{equation}
P(N) = \frac{1}{\pi(1+|N|^2)}
\label{eq:cauchy}
\end{equation}

Finally, we use the Student's 
$t$ distribution characterized by $n$ (which is sometimes referred to as ``degrees of freedom'') and is given by
\begin{equation}
P(N) = \frac{\Gamma[(n+1)/2]}{\sqrt{\pi n}\Gamma(n/2)(1+|N|^2/n)^{(n+1)/2}}
\label{eq:student}
\end{equation}
\noindent For $n=1$, the Student's $t$ distribution is same as the  Cauchy distribution, and is equal to Gaussian distribution for $n=\infty$. For our analysis, we vary $n$ from 2 to 2000. Note that the Students-$t$ distribution for the error residuals can be obtained by modeling  the error in  systematic errors as a gamma distribution~\citep{Cowan18}.

In addition to comparing the  error distributions  to the PDFs in Eqs.~\ref{eq:gauss},~\ref{eq:laplace},~\ref{eq:cauchy}, ~\ref{eq:student}, which mainly depend on $|N|$, we also compare to these distributions, after replacing $N$ by $N/S$, where $S$ is an arbitrary scale factor, which we vary  from 0.001 to 2.5 in steps of size 0.01. 

The comparison is done using the one-sample unbinned Kolmogorov-Smirnov (K-S) test~\cite{astroML}.  The K-S test is based on the $D$-statistic, which measures the maximum distance between two cumulative distributions. The K-S test is widely used in both astrophysics and particle physics, for comparison of a dataset to a wide range of probability distributions,  as it is agnostic to the distribution against which it  is been tested, and does not depend on the size  of the sample. Furthermore, critical values based upon the $D$-statistic have been calculated in the literature and can be easily computed for any value of $D$.  This test is also invariant to reparameterization of the data. The one-sample K-S test can therefore   serve as a goodness-of-fit test. Although some concerns have been raised regarding the incorrect usage of K-S test in astrophysics literature, as well as other caveats and limitations of this test~\cite{Babu}, these do not apply  in our case, and hence we use the K-S test to evaluate the compatibility of the error residuals with various distributions.
In this case, the two distributions are the error histograms and the parent PDF to which it is compared. From the $D$ statistic, the K-S test also provides a $p$-value, whose analytic formula can be found in any statistic work~\cite{astroML,RatraD2}. For this work, we have used the {\tt scipy} module in {\tt Python} for the computations. Higher the $p$-value, more similar are the two distributions, whereas a low $p$-value indicates an inconsistency between the  distributions. Our results for comparison with all the four distributions are summarized in Table~\ref{tab:ks}.

%changed not a good fit to not the best fit
We find that for all  three estimates, the Gaussian distribution is not the best fit, unless the scale factor is different from unity. The data are much more consistent with Cauchy or Student's $t$ distribution. However, none of the $p$-values  for the  Gaussian distribution  are small enough to reject the null hypothesis.

\begin{table}
\caption{Probabilities from K-S test for various distributions using the observed neutron lifetime measurements.}
\begin{threeparttable}[t]
\begin{tabular}{cccc} 
\hline
%&\multicolumn{2}{c}{Truncated 13}& \\ \hline  

Distribution&$S$\tnote{a}&$p$\tnote{b}&$n$\tnote{c} \\ \hline
\multicolumn{4}{l}{\textbf{Median} ($\tau_{med}$)} \\ 
\multicolumn{1}{l}{Gaussian}&1&$0.299$& \\
&$1.317$&$0.875$& \\
\multicolumn{1}{l}{Laplacian}&1&$0.771$&  \\
&$1.214$&$0.983$&\\
\multicolumn{1}{l}{Cauchy}&1&$0.878$& \\
&$0.786$&$0.997$& \\
\multicolumn{1}{l}{Student's $t$}&1&$0.954$&$2$ \\
&$1.021$&$0.966$&$2$ \\
\multicolumn{4}{l}{\textbf{Weighted Mean ($\tau_{wm+}$)}} \\ 
\multicolumn{1}{l}{Gaussian}&1&$0.327$& \\
&$1.378$&$0.974$& \\
\multicolumn{1}{l}{Laplacian}&1&$0.691$&  \\
&$1.304$&$0.996$&\\
\multicolumn{1}{l}{Cauchy}&1&$0.908$& \\
&$0.817$&$0.982$& \\
\multicolumn{1}{l}{Student's  $t$}&1&$0.928$&$2$ \\
&$1.091$&$0.989$&$2$ \\
\multicolumn{4}{l}{\textbf{Weighted Mean ($\tau_{wm-}$) }} \\ 
\multicolumn{1}{l}{Gaussian}&1&$0.186$& \\
&$1.562$&$0.958$& \\
\multicolumn{1}{l}{Laplacian}&1&$0.556$&  \\
&$1.428$&$0.996$&\\
\multicolumn{1}{l}{Cauchy}&1&$0.925$& \\
&$0.85$&$0.980$& \\
\multicolumn{1}{l}{Student's $t$}&1&$0.267$&$2$ \\
&$1.201$&$0.987$&$2$ \\

\hline
\end{tabular}
\begin{tablenotes}
\item[a] \footnotesize{The scale factor (other than 1)  which maximizes $p$}
\item[b] \footnotesize{$p$-value that the data is derived from the PDF}
\item[c] \footnotesize{The value $n$ in the students $t$-distribution}
\end{tablenotes}

\end{threeparttable}
\label{tab:ks}
\end{table}

\section{Discrepancy between beam and bottle measurements}
\label{sec:disc}
We now quantify the significance of the discrepancy between beam and bottle-based experiments using central estimates based on  the  median statistics. We do this analysis using three different combinations of datasets for beam and bottle based experiments. A summary of these comparisons can be found in Table~\ref{tab:4}.

We first use the same datapoints  as in F18~\citep{Fornal}, who argued for a $4.4\sigma$ discrepancy. 
We obtain a median estimate using the same bottle-based experiments  considered in F18~\cite{Mampe,Serebrov05,Pichlmaier10,Steyerl2,Arzumanov15}, and compare the same  with the beam-based experiments therein~\citep{Byrne96,Yue13}. The median lifetime of the five bottle-based experiments along with the $1\sigma$ median error bar is given by $880.7 \pm 1.3$ seconds. The corresponding lifetime for the two beam-based experiments considered in F18 is $888.45$ seconds. Since, it is not possible to obtain a median error estimate with just two measurements, we do not quote its $1\sigma$ median uncertainty. The results do not change even after including the two additional bottle-based measurements~\citep{Serebrov17,Pattie} not used for their average. Therefore considering the median statistics estimates, the discrepancy is about $6\sigma$.

If we do this comparison by including all the measurements in Table~\ref{tab:4}, the median lifetime for all the  bottle-based experiments is equal to  $880.7 \pm 1.2$ seconds. The corresponding number for all the beam-based experiments is $888.45 \pm 1.65$ seconds. Therefore, comparing the median estimates between the beam and bottle-based  measurements amounts to a 3.79$\sigma$ discrepancy. 

If we then redo this comparison for a subset of all measurements in Table~\ref{tab:1}, having total error less than 10 seconds, the median central estimate for all bottle-based experiments is  $880.2 \pm 1.1$ secs. Since the total number of
beam-based measurements in Table~\ref{tab:1} is  a very small number (three), we only can obtain a central estimate, which is equal to 889.2 seconds. Therefore, the total discrepancy is about $8.2\sigma$. 

Hence, we infer that the discrepancy  between beam and bottle-based measurements persists, even when  median statistics is used for the central estimate of the neutron lifetime.

\begin{table*}
\begin{tabular}{|c|c|c|c|}
\hline
Dataset & $\tau_N$ (bottle-based) &  $\tau_N$ (beam-based) &  Discrepancy \\
& (secs) & (secs) & \\
\hline
F18~\citep{Fornal} & $880.7 \pm 1.3$ & $888.45$ & $6\sigma$ \\ \hline
Data from Table~\ref{tab:1} & $880.7 \pm 1.2$ & $888.45 \pm 1.65$ & $3.79\sigma$ \\ \hline
Data from Table 1   & $880.2 \pm 1.1$ & 889.2 & 8.2$\sigma$ \\ 
having errors $<$ 10 secs & & & \\ \hline
\end{tabular}
\caption{Summary of the significance of the discrepancies between beam and bottle-based measurements using median statistics. The first column refers to the datasets used. The second and third columns contain the median statistics estimate of the neutron lifetime($\tau_N$) using bottle and beam-based measurements respectively, using $1\sigma$ median error bars obtained using the procedure in Ref.~\cite{Gott}. The last column indicates the statistical significance of the discrepancy.} 
\label{tab:4}
\end{table*}

\section{Conclusions}
\label{sec:conclusions}

There has been a long-standing discrepancy in literature related to the neutron lifetime measurements between the two different techniques, viz. bottle and beam-based methods. As of 2019, the current discrepancy is about $4\sigma$~\cite{Fornal}. 
To get some insight into these issues, we carried out an extensive meta-analysis of the vetted neutron lifetime measurements compiled in literature.
We first use  a compilation of 19 measurements of the neutron lifetime and their corresponding errors listed in the  2019 edition of PDG~\citep{PDG} (cf. Table~\ref{tab:1}), in order to ascertain the non-Gaussianity of the residuals and to  obtain a central estimate.  The error distributions were analyzed in the same way as previously done for a variety of  astrophysical measurements  by Ratra et al~\cite{RatraD2,RatraGC,Ratratheta}. For this purpose, the central estimate  was obtained  using both the weighted mean (with and without correlations)  as well as the median value. The median estimate does not incorporate the errors in the neutron lifetime. We then fit these residuals to four distributions, viz. Gaussian, Laplace, Cauchy, and Student's $t$ distribution.
The resulting fits are tabulated in Table~\ref{tab:ks}.

We conclude from  these observations,  
that none of the $p$-values (obtained
using all the three central estimates) are small enough to reject the Gaussian distribution for the error residuals. 
%We also checked using simulations that the $p$-values for a sample of 19 measurements have a large dynamic range and are sometimes close to 0, even for the null-hypothesis. 
However,  the Student's $t$ and  Cauchy distributions provide a more robust fit than the Gaussian distribution.

Therefore, more data points  are necessary to robustly determine if the error residuals are consistent with a Gaussian distribution. Nevertheless, it  would be useful exercise to 
obtain the central estimate of the neutron lifetime with median statistic, and to check if  the discrepancy between beam and bottle-based measurements persists using median statistics.
This median value along with 1$\sigma$ error bars using the 19 measurements, which we obtain  is given by $881.5 \pm 0.47$ seconds. This estimate is complementary to the PDG-based result obtained using weighted mean statistic, which includes the addition of an  ad-hoc scale factor.
\rthis{This value can be used as an alternate estimate of the observed neutron lifetime, and used for comparison with the theoretical estimate, which is currently between 875.3 and 891.2 seconds within $3\sigma$~\cite{Fornal}.} 
Furthermore, this median value provides an alternate central estimate of the neutron lifetime, which can be used for comparison with theoretical estimates.

We then used  the median estimate to evaluate the statistical significance of the discrepancy between beam and bottle-based measurements. When we use the same measurements as in F18, the discrepancy exacerbates  to 6$\sigma$. If we consider all the measurements in Table~\ref{tab:1}, the discrepancy becomes $3.8\sigma$ ($8.2\sigma$), depending on whether we include (exclude) measurements in this, with total error less than 10 seconds.

\begin{acknowledgements}
We are grateful to Tomasso Dorigo for his nice blog article about the F18 paper, which brought our attention to this problem. We also thank Bharat Ratra for explaining in detail the methodology used in P18 and also in his earlier works.
\end{acknowledgements}

\bibliography{main}

\begin{thebibliography}{49}
\expandafter\ifx\csname natexlab\endcsname\relax\def\natexlab#1{#1}\fi
\expandafter\ifx\csname bibnamefont\endcsname\relax
  \def\bibnamefont#1{#1}\fi
\expandafter\ifx\csname bibfnamefont\endcsname\relax
  \def\bibfnamefont#1{#1}\fi
\expandafter\ifx\csname citenamefont\endcsname\relax
  \def\citenamefont#1{#1}\fi
\expandafter\ifx\csname url\endcsname\relax
  \def\url#1{\texttt{#1}}\fi
\expandafter\ifx\csname urlprefix\endcsname\relax\def\urlprefix{URL }\fi
\providecommand{\bibinfo}[2]{#2}
\providecommand{\eprint}[2][]{\url{#2}}

\bibitem[{\citenamefont{Wietfeldt and Greene}(2011)}]{neutronlifetime11}
\bibinfo{author}{\bibfnamefont{F.~E.} \bibnamefont{Wietfeldt}}
  \bibnamefont{and} \bibinfo{author}{\bibfnamefont{G.~L.}
  \bibnamefont{Greene}}, \bibinfo{journal}{Reviews of Modern Physics}
  \textbf{\bibinfo{volume}{83}}, \bibinfo{pages}{1173} (\bibinfo{year}{2011}).

\bibitem[{\citenamefont{Wietfeldt}(2014)}]{neutronlifetime14}
\bibinfo{author}{\bibfnamefont{F.~E.} \bibnamefont{Wietfeldt}}, in
  \emph{\bibinfo{booktitle}{{8th International Workshop on the CKM Unitarity
  Triangle (CKM 2014) Vienna, Austria, September 8-12, 2014}}}
  (\bibinfo{year}{2014}), \eprint{1411.3687}.

\bibitem[{\citenamefont{Tanabashi et~al.}(2018)}]{PDG}
\bibinfo{author}{\bibfnamefont{M.}~\bibnamefont{Tanabashi}}
  \bibnamefont{et~al.} (\bibinfo{collaboration}{Particle Data Group}),
  \bibinfo{journal}{Phys. Rev.} \textbf{\bibinfo{volume}{D98}},
  \bibinfo{pages}{030001} (\bibinfo{year}{2018}).

\bibitem[{\citenamefont{{Press} et~al.}(1992)\citenamefont{{Press},
  {Teukolsky}, {Vetterling}, and {Flannery}}}]{NR}
\bibinfo{author}{\bibfnamefont{W.~H.} \bibnamefont{{Press}}},
  \bibinfo{author}{\bibfnamefont{S.~A.} \bibnamefont{{Teukolsky}}},
  \bibinfo{author}{\bibfnamefont{W.~T.} \bibnamefont{{Vetterling}}},
  \bibnamefont{and} \bibinfo{author}{\bibfnamefont{B.~P.}
  \bibnamefont{{Flannery}}}, \emph{\bibinfo{title}{{Numerical recipes in C. The
  art of scientific computing}}} (\bibinfo{year}{1992}).

\bibitem[{\citenamefont{{Ganguly} and {Desai}}(2017)}]{Ganguly}
\bibinfo{author}{\bibfnamefont{S.}~\bibnamefont{{Ganguly}}} \bibnamefont{and}
  \bibinfo{author}{\bibfnamefont{S.}~\bibnamefont{{Desai}}},
  \bibinfo{journal}{Astroparticle Physics} \textbf{\bibinfo{volume}{94}},
  \bibinfo{pages}{17} (\bibinfo{year}{2017}), \eprint{1706.01202}.

\bibitem[{\citenamefont{Fornal and Grinstein}(2018)}]{Fornal}
\bibinfo{author}{\bibfnamefont{B.}~\bibnamefont{Fornal}} \bibnamefont{and}
  \bibinfo{author}{\bibfnamefont{B.}~\bibnamefont{Grinstein}},
  \bibinfo{journal}{Physical Review Letters} \textbf{\bibinfo{volume}{120}},
  \bibinfo{pages}{191801} (\bibinfo{year}{2018}).

\bibitem[{\citenamefont{Czarnecki et~al.}(2018)\citenamefont{Czarnecki,
  Marciano, and Sirlin}}]{Marciano}
\bibinfo{author}{\bibfnamefont{A.}~\bibnamefont{Czarnecki}},
  \bibinfo{author}{\bibfnamefont{W.~J.} \bibnamefont{Marciano}},
  \bibnamefont{and} \bibinfo{author}{\bibfnamefont{A.}~\bibnamefont{Sirlin}},
  \bibinfo{journal}{Phys. Rev. Lett.} \textbf{\bibinfo{volume}{120}},
  \bibinfo{pages}{202002} (\bibinfo{year}{2018}), \eprint{1802.01804}.

\bibitem[{\citenamefont{Greene and Geltenbort}(2016)}]{Greene}
\bibinfo{author}{\bibfnamefont{G.~L.} \bibnamefont{Greene}} \bibnamefont{and}
  \bibinfo{author}{\bibfnamefont{P.}~\bibnamefont{Geltenbort}},
  \bibinfo{journal}{Scientific American} \textbf{\bibinfo{volume}{314}},
  \bibinfo{pages}{36} (\bibinfo{year}{2016}).

\bibitem[{\citenamefont{Byrne et~al.}(1996)\citenamefont{Byrne, Dawber, Habeck,
  Smidt, Spain, and Williams}}]{Byrne96}
\bibinfo{author}{\bibfnamefont{J.}~\bibnamefont{Byrne}},
  \bibinfo{author}{\bibfnamefont{P.}~\bibnamefont{Dawber}},
  \bibinfo{author}{\bibfnamefont{C.}~\bibnamefont{Habeck}},
  \bibinfo{author}{\bibfnamefont{S.}~\bibnamefont{Smidt}},
  \bibinfo{author}{\bibfnamefont{J.}~\bibnamefont{Spain}}, \bibnamefont{and}
  \bibinfo{author}{\bibfnamefont{A.}~\bibnamefont{Williams}},
  \bibinfo{journal}{EPL (Europhysics Letters)} \textbf{\bibinfo{volume}{33}},
  \bibinfo{pages}{187} (\bibinfo{year}{1996}).

\bibitem[{\citenamefont{Yue et~al.}(2013)\citenamefont{Yue, Dewey, Gilliam,
  Greene, Laptev, Nico, Snow, and Wietfeldt}}]{Yue13}
\bibinfo{author}{\bibfnamefont{A.}~\bibnamefont{Yue}},
  \bibinfo{author}{\bibfnamefont{M.}~\bibnamefont{Dewey}},
  \bibinfo{author}{\bibfnamefont{D.}~\bibnamefont{Gilliam}},
  \bibinfo{author}{\bibfnamefont{G.}~\bibnamefont{Greene}},
  \bibinfo{author}{\bibfnamefont{A.}~\bibnamefont{Laptev}},
  \bibinfo{author}{\bibfnamefont{J.}~\bibnamefont{Nico}},
  \bibinfo{author}{\bibfnamefont{W.~M.} \bibnamefont{Snow}}, \bibnamefont{and}
  \bibinfo{author}{\bibfnamefont{F.}~\bibnamefont{Wietfeldt}},
  \bibinfo{journal}{Physical review letters} \textbf{\bibinfo{volume}{111}},
  \bibinfo{pages}{222501} (\bibinfo{year}{2013}).

\bibitem[{\citenamefont{Mampe et~al.}(1993)\citenamefont{Mampe, Bondarenko,
  Morozov, Panin, and Fomin}}]{Mampe}
\bibinfo{author}{\bibfnamefont{W.}~\bibnamefont{Mampe}},
  \bibinfo{author}{\bibfnamefont{L.}~\bibnamefont{Bondarenko}},
  \bibinfo{author}{\bibfnamefont{V.}~\bibnamefont{Morozov}},
  \bibinfo{author}{\bibfnamefont{Y.~N.} \bibnamefont{Panin}}, \bibnamefont{and}
  \bibinfo{author}{\bibfnamefont{A.}~\bibnamefont{Fomin}},
  \bibinfo{journal}{JETP Letters} \textbf{\bibinfo{volume}{57}},
  \bibinfo{pages}{82} (\bibinfo{year}{1993}).

\bibitem[{\citenamefont{Serebrov et~al.}(2005)\citenamefont{Serebrov, Varlamov,
  Kharitonov, Fomin, Pokotilovski, Geltenbort, Butterworth, Krasnoschekova,
  Lasakov, Tal'daev et~al.}}]{Serebrov05}
\bibinfo{author}{\bibfnamefont{A.}~\bibnamefont{Serebrov}},
  \bibinfo{author}{\bibfnamefont{V.}~\bibnamefont{Varlamov}},
  \bibinfo{author}{\bibfnamefont{A.}~\bibnamefont{Kharitonov}},
  \bibinfo{author}{\bibfnamefont{A.}~\bibnamefont{Fomin}},
  \bibinfo{author}{\bibfnamefont{Y.}~\bibnamefont{Pokotilovski}},
  \bibinfo{author}{\bibfnamefont{P.}~\bibnamefont{Geltenbort}},
  \bibinfo{author}{\bibfnamefont{J.}~\bibnamefont{Butterworth}},
  \bibinfo{author}{\bibfnamefont{I.}~\bibnamefont{Krasnoschekova}},
  \bibinfo{author}{\bibfnamefont{M.}~\bibnamefont{Lasakov}},
  \bibinfo{author}{\bibfnamefont{R.}~\bibnamefont{Tal'daev}},
  \bibnamefont{et~al.}, \bibinfo{journal}{Physics Letters B}
  \textbf{\bibinfo{volume}{605}}, \bibinfo{pages}{72} (\bibinfo{year}{2005}).

\bibitem[{\citenamefont{Pichlmaier et~al.}(2010)\citenamefont{Pichlmaier,
  Varlamov, Schreckenbach, and Geltenbort}}]{Pichlmaier10}
\bibinfo{author}{\bibfnamefont{A.}~\bibnamefont{Pichlmaier}},
  \bibinfo{author}{\bibfnamefont{V.}~\bibnamefont{Varlamov}},
  \bibinfo{author}{\bibfnamefont{K.}~\bibnamefont{Schreckenbach}},
  \bibnamefont{and}
  \bibinfo{author}{\bibfnamefont{P.}~\bibnamefont{Geltenbort}},
  \bibinfo{journal}{Physics Letters B} \textbf{\bibinfo{volume}{693}},
  \bibinfo{pages}{221} (\bibinfo{year}{2010}).

\bibitem[{\citenamefont{Steyerl et~al.}(2012)\citenamefont{Steyerl, Pendlebury,
  Kaufman, Malik, and Desai}}]{Steyerl2}
\bibinfo{author}{\bibfnamefont{A.}~\bibnamefont{Steyerl}},
  \bibinfo{author}{\bibfnamefont{J.}~\bibnamefont{Pendlebury}},
  \bibinfo{author}{\bibfnamefont{C.}~\bibnamefont{Kaufman}},
  \bibinfo{author}{\bibfnamefont{S.~S.} \bibnamefont{Malik}}, \bibnamefont{and}
  \bibinfo{author}{\bibfnamefont{A.}~\bibnamefont{Desai}},
  \bibinfo{journal}{Physical Review C} \textbf{\bibinfo{volume}{85}},
  \bibinfo{pages}{065503} (\bibinfo{year}{2012}).

\bibitem[{\citenamefont{Arzumanov et~al.}(2015)\citenamefont{Arzumanov,
  Bondarenko, Chernyavsky, Geltenbort, Morozov, Nesvizhevsky, Panin, and
  Strepetov}}]{Arzumanov15}
\bibinfo{author}{\bibfnamefont{S.}~\bibnamefont{Arzumanov}},
  \bibinfo{author}{\bibfnamefont{L.}~\bibnamefont{Bondarenko}},
  \bibinfo{author}{\bibfnamefont{S.}~\bibnamefont{Chernyavsky}},
  \bibinfo{author}{\bibfnamefont{P.}~\bibnamefont{Geltenbort}},
  \bibinfo{author}{\bibfnamefont{V.}~\bibnamefont{Morozov}},
  \bibinfo{author}{\bibfnamefont{V.}~\bibnamefont{Nesvizhevsky}},
  \bibinfo{author}{\bibfnamefont{Y.}~\bibnamefont{Panin}}, \bibnamefont{and}
  \bibinfo{author}{\bibfnamefont{A.}~\bibnamefont{Strepetov}},
  \bibinfo{journal}{Physics Letters B} \textbf{\bibinfo{volume}{745}},
  \bibinfo{pages}{79} (\bibinfo{year}{2015}).

\bibitem[{\citenamefont{{Gott} et~al.}(2001)\citenamefont{{Gott}, {Vogeley},
  {Podariu}, and {Ratra}}}]{Gott}
\bibinfo{author}{\bibfnamefont{J.~R.} \bibnamefont{{Gott}},
  \bibfnamefont{III}}, \bibinfo{author}{\bibfnamefont{M.~S.}
  \bibnamefont{{Vogeley}}},
  \bibinfo{author}{\bibfnamefont{S.}~\bibnamefont{{Podariu}}},
  \bibnamefont{and} \bibinfo{author}{\bibfnamefont{B.}~\bibnamefont{{Ratra}}},
  \bibinfo{journal}{\apj} \textbf{\bibinfo{volume}{549}}, \bibinfo{pages}{1}
  (\bibinfo{year}{2001}), \eprint{astro-ph/0006103}.

\bibitem[{\citenamefont{{Bethapudi} and {Desai}}(2017)}]{Bethapudi}
\bibinfo{author}{\bibfnamefont{S.}~\bibnamefont{{Bethapudi}}} \bibnamefont{and}
  \bibinfo{author}{\bibfnamefont{S.}~\bibnamefont{{Desai}}},
  \bibinfo{journal}{European Physical Journal Plus}
  \textbf{\bibinfo{volume}{132}}, \bibinfo{eid}{78} (\bibinfo{year}{2017}),
  \eprint{1701.01789}.

\bibitem[{\citenamefont{{Chen} and {Ratra}}(2003)}]{Chen03}
\bibinfo{author}{\bibfnamefont{G.}~\bibnamefont{{Chen}}} \bibnamefont{and}
  \bibinfo{author}{\bibfnamefont{B.}~\bibnamefont{{Ratra}}},
  \bibinfo{journal}{\pasp} \textbf{\bibinfo{volume}{115}},
  \bibinfo{pages}{1143} (\bibinfo{year}{2003}), \eprint{astro-ph/0302002}.

\bibitem[{\citenamefont{{Chen} and {Ratra}}(2011)}]{Chen}
\bibinfo{author}{\bibfnamefont{G.}~\bibnamefont{{Chen}}} \bibnamefont{and}
  \bibinfo{author}{\bibfnamefont{B.}~\bibnamefont{{Ratra}}},
  \bibinfo{journal}{\pasp} \textbf{\bibinfo{volume}{123}},
  \bibinfo{pages}{1127} (\bibinfo{year}{2011}), \eprint{1105.5206}.

\bibitem[{\citenamefont{{Chen} et~al.}(2003)\citenamefont{{Chen}, {Gott}, and
  {Ratra}}}]{Ratra03}
\bibinfo{author}{\bibfnamefont{G.}~\bibnamefont{{Chen}}},
  \bibinfo{author}{\bibfnamefont{J.~R.} \bibnamefont{{Gott}},
  \bibfnamefont{III}}, \bibnamefont{and}
  \bibinfo{author}{\bibfnamefont{B.}~\bibnamefont{{Ratra}}},
  \bibinfo{journal}{\pasp} \textbf{\bibinfo{volume}{115}},
  \bibinfo{pages}{1269} (\bibinfo{year}{2003}), \eprint{astro-ph/0308099}.

\bibitem[{\citenamefont{{Crandall} et~al.}(2015)\citenamefont{{Crandall},
  {Houston}, and {Ratra}}}]{Ratra15}
\bibinfo{author}{\bibfnamefont{S.}~\bibnamefont{{Crandall}}},
  \bibinfo{author}{\bibfnamefont{S.}~\bibnamefont{{Houston}}},
  \bibnamefont{and} \bibinfo{author}{\bibfnamefont{B.}~\bibnamefont{{Ratra}}},
  \bibinfo{journal}{Modern Physics Letters A} \textbf{\bibinfo{volume}{30}},
  \bibinfo{eid}{1550123} (\bibinfo{year}{2015}), \eprint{1409.7332}.

\bibitem[{\citenamefont{{Crandall} and {Ratra}}(2015)}]{RatraLMC}
\bibinfo{author}{\bibfnamefont{S.}~\bibnamefont{{Crandall}}} \bibnamefont{and}
  \bibinfo{author}{\bibfnamefont{B.}~\bibnamefont{{Ratra}}},
  \bibinfo{journal}{\apj} \textbf{\bibinfo{volume}{815}}, \bibinfo{eid}{87}
  (\bibinfo{year}{2015}), \eprint{1507.07940}.

\bibitem[{\citenamefont{{Crandall} and {Ratra}}(2014)}]{Crandall}
\bibinfo{author}{\bibfnamefont{S.}~\bibnamefont{{Crandall}}} \bibnamefont{and}
  \bibinfo{author}{\bibfnamefont{B.}~\bibnamefont{{Ratra}}},
  \bibinfo{journal}{Physics Letters B} \textbf{\bibinfo{volume}{732}},
  \bibinfo{pages}{330} (\bibinfo{year}{2014}), \eprint{1311.0840}.

\bibitem[{\citenamefont{{Rajan} and {Desai}}(2018)}]{Rajan}
\bibinfo{author}{\bibfnamefont{A.}~\bibnamefont{{Rajan}}} \bibnamefont{and}
  \bibinfo{author}{\bibfnamefont{S.}~\bibnamefont{{Desai}}},
  \bibinfo{journal}{European Physical Journal Plus}
  \textbf{\bibinfo{volume}{133}}, \bibinfo{eid}{107} (\bibinfo{year}{2018}),
  \eprint{1710.06624}.

\bibitem[{\citenamefont{{Penton} et~al.}(2018)\citenamefont{{Penton}, {Peyton},
  {Zahoor}, and {Ratra}}}]{RatraD2}
\bibinfo{author}{\bibfnamefont{J.}~\bibnamefont{{Penton}}},
  \bibinfo{author}{\bibfnamefont{J.}~\bibnamefont{{Peyton}}},
  \bibinfo{author}{\bibfnamefont{A.}~\bibnamefont{{Zahoor}}}, \bibnamefont{and}
  \bibinfo{author}{\bibfnamefont{B.}~\bibnamefont{{Ratra}}},
  \bibinfo{journal}{\pasp} \textbf{\bibinfo{volume}{130}},
  \bibinfo{pages}{114001} (\bibinfo{year}{2018}), \eprint{1808.01490}.

\bibitem[{\citenamefont{{Camarillo}
  et~al.}(2018{\natexlab{a}})\citenamefont{{Camarillo}, {Dredger}, and
  {Ratra}}}]{Ratratheta}
\bibinfo{author}{\bibfnamefont{T.}~\bibnamefont{{Camarillo}}},
  \bibinfo{author}{\bibfnamefont{P.}~\bibnamefont{{Dredger}}},
  \bibnamefont{and} \bibinfo{author}{\bibfnamefont{B.}~\bibnamefont{{Ratra}}},
  \bibinfo{journal}{\apss} \textbf{\bibinfo{volume}{363}}, \bibinfo{eid}{268}
  (\bibinfo{year}{2018}{\natexlab{a}}), \eprint{1805.01917}.

\bibitem[{\citenamefont{{Zhang}}(2017)}]{Zhang}
\bibinfo{author}{\bibfnamefont{J.}~\bibnamefont{{Zhang}}},
  \bibinfo{journal}{\mnras} \textbf{\bibinfo{volume}{468}},
  \bibinfo{pages}{5014} (\bibinfo{year}{2017}).

\bibitem[{\citenamefont{{Camarillo}
  et~al.}(2018{\natexlab{b}})\citenamefont{{Camarillo}, {Mathur}, {Mitchell},
  and {Ratra}}}]{RatraGC}
\bibinfo{author}{\bibfnamefont{T.}~\bibnamefont{{Camarillo}}},
  \bibinfo{author}{\bibfnamefont{V.}~\bibnamefont{{Mathur}}},
  \bibinfo{author}{\bibfnamefont{T.}~\bibnamefont{{Mitchell}}},
  \bibnamefont{and} \bibinfo{author}{\bibfnamefont{B.}~\bibnamefont{{Ratra}}},
  \bibinfo{journal}{\pasp} \textbf{\bibinfo{volume}{130}},
  \bibinfo{pages}{024101} (\bibinfo{year}{2018}{\natexlab{b}}),
  \eprint{1708.01310}.

\bibitem[{\citenamefont{{Bailey}}(2017)}]{Bailey}
\bibinfo{author}{\bibfnamefont{D.~C.} \bibnamefont{{Bailey}}},
  \bibinfo{journal}{Royal Society Open Science} \textbf{\bibinfo{volume}{4}},
  \bibinfo{eid}{160600} (\bibinfo{year}{2017}), \eprint{1612.00778}.

\bibitem[{\citenamefont{Cowan}(2019)}]{Cowan18}
\bibinfo{author}{\bibfnamefont{G.}~\bibnamefont{Cowan}}, \bibinfo{journal}{Eur.
  Phys. J.} \textbf{\bibinfo{volume}{C79}}, \bibinfo{pages}{133}
  (\bibinfo{year}{2019}), \eprint{1809.05778}.

\bibitem[{\citenamefont{Leung et~al.}(2016)\citenamefont{Leung, Geltenbort,
  Ivanov, Rosenau, and Zimmer}}]{Leung}
\bibinfo{author}{\bibfnamefont{K.~K.~H.} \bibnamefont{Leung}},
  \bibinfo{author}{\bibfnamefont{P.}~\bibnamefont{Geltenbort}},
  \bibinfo{author}{\bibfnamefont{S.}~\bibnamefont{Ivanov}},
  \bibinfo{author}{\bibfnamefont{F.}~\bibnamefont{Rosenau}}, \bibnamefont{and}
  \bibinfo{author}{\bibfnamefont{O.}~\bibnamefont{Zimmer}},
  \bibinfo{journal}{Phys. Rev.} \textbf{\bibinfo{volume}{C94}},
  \bibinfo{pages}{045502} (\bibinfo{year}{2016}), \eprint{1606.00929}.

\bibitem[{\citenamefont{{Salvati} et~al.}(2016)\citenamefont{{Salvati},
  {Pagano}, {Consiglio}, and {Melchiorri}}}]{Salvati}
\bibinfo{author}{\bibfnamefont{L.}~\bibnamefont{{Salvati}}},
  \bibinfo{author}{\bibfnamefont{L.}~\bibnamefont{{Pagano}}},
  \bibinfo{author}{\bibfnamefont{R.}~\bibnamefont{{Consiglio}}},
  \bibnamefont{and}
  \bibinfo{author}{\bibfnamefont{A.}~\bibnamefont{{Melchiorri}}},
  \bibinfo{journal}{\jcap} \textbf{\bibinfo{volume}{3}}, \bibinfo{eid}{055}
  (\bibinfo{year}{2016}), \eprint{1507.07243}.

\bibitem[{\citenamefont{Woodruff}(1952)}]{Woodruff}
\bibinfo{author}{\bibfnamefont{R.~S.} \bibnamefont{Woodruff}},
  \bibinfo{journal}{Journal of the American Statistical Association}
  \textbf{\bibinfo{volume}{47}}, \bibinfo{pages}{635} (\bibinfo{year}{1952}).

\bibitem[{\citenamefont{Maritz and Jarrett}(1978)}]{Maritz}
\bibinfo{author}{\bibfnamefont{J.}~\bibnamefont{Maritz}} \bibnamefont{and}
  \bibinfo{author}{\bibfnamefont{R.}~\bibnamefont{Jarrett}},
  \bibinfo{journal}{Journal of the American Statistical Association}
  \textbf{\bibinfo{volume}{73}}, \bibinfo{pages}{194} (\bibinfo{year}{1978}).

\bibitem[{\citenamefont{Price and Bonett}(2001)}]{Price}
\bibinfo{author}{\bibfnamefont{R.~M.} \bibnamefont{Price}} \bibnamefont{and}
  \bibinfo{author}{\bibfnamefont{D.~G.} \bibnamefont{Bonett}},
  \bibinfo{journal}{Journal of Statistical Computation and Simulation}
  \textbf{\bibinfo{volume}{68}}, \bibinfo{pages}{295} (\bibinfo{year}{2001}).

\bibitem[{\citenamefont{{Bevington} and {Robinson}}(1992)}]{Bevington}
\bibinfo{author}{\bibfnamefont{P.~R.} \bibnamefont{{Bevington}}}
  \bibnamefont{and} \bibinfo{author}{\bibfnamefont{D.~K.}
  \bibnamefont{{Robinson}}}, \emph{\bibinfo{title}{{Data reduction and error
  analysis for the physical sciences}}} (\bibinfo{year}{1992}).

\bibitem[{\citenamefont{Ezhov et~al.}(2018)\citenamefont{Ezhov, Andreev, Ban,
  Bazarov, Geltenbort, Glushkov, Knyazkov, Kovrizhnykh, Krygin, Naviliat-Cuncic
  et~al.}}]{Ezhov18}
\bibinfo{author}{\bibfnamefont{V.}~\bibnamefont{Ezhov}},
  \bibinfo{author}{\bibfnamefont{A.}~\bibnamefont{Andreev}},
  \bibinfo{author}{\bibfnamefont{G.}~\bibnamefont{Ban}},
  \bibinfo{author}{\bibfnamefont{B.}~\bibnamefont{Bazarov}},
  \bibinfo{author}{\bibfnamefont{P.}~\bibnamefont{Geltenbort}},
  \bibinfo{author}{\bibfnamefont{A.}~\bibnamefont{Glushkov}},
  \bibinfo{author}{\bibfnamefont{V.}~\bibnamefont{Knyazkov}},
  \bibinfo{author}{\bibfnamefont{N.~A.} \bibnamefont{Kovrizhnykh}},
  \bibinfo{author}{\bibfnamefont{G.}~\bibnamefont{Krygin}},
  \bibinfo{author}{\bibfnamefont{O.}~\bibnamefont{Naviliat-Cuncic}},
  \bibnamefont{et~al.}, \bibinfo{journal}{JETP Letters}
  \textbf{\bibinfo{volume}{107}}, \bibinfo{pages}{671} (\bibinfo{year}{2018}).

\bibitem[{\citenamefont{Serebrov et~al.}(2018)}]{Serebrov17}
\bibinfo{author}{\bibfnamefont{A.~P.} \bibnamefont{Serebrov}}
  \bibnamefont{et~al.}, \bibinfo{journal}{Phys. Rev.}
  \textbf{\bibinfo{volume}{C97}}, \bibinfo{pages}{055503}
  (\bibinfo{year}{2018}), \eprint{1712.05663}.

\bibitem[{\citenamefont{Pattie et~al.}(2018)}]{Pattie}
\bibinfo{author}{\bibfnamefont{R.~W.} \bibnamefont{Pattie}, \bibfnamefont{Jr.}}
  \bibnamefont{et~al.}, \bibinfo{journal}{Science}
  \textbf{\bibinfo{volume}{360}}, \bibinfo{pages}{627} (\bibinfo{year}{2018}),
  \eprint{1707.01817}.

\bibitem[{\citenamefont{Alfimenkov et~al.}(1990)\citenamefont{Alfimenkov,
  Varlamov, Vasil’ev, Gudkov, Lushchikov, Nesvizhevskii, Serebrov, Strelkov,
  Sumbaev, Tal’daev et~al.}}]{Alfimenkov90}
\bibinfo{author}{\bibfnamefont{V.}~\bibnamefont{Alfimenkov}},
  \bibinfo{author}{\bibfnamefont{V.}~\bibnamefont{Varlamov}},
  \bibinfo{author}{\bibfnamefont{A.}~\bibnamefont{Vasil’ev}},
  \bibinfo{author}{\bibfnamefont{V.}~\bibnamefont{Gudkov}},
  \bibinfo{author}{\bibfnamefont{V.}~\bibnamefont{Lushchikov}},
  \bibinfo{author}{\bibfnamefont{V.}~\bibnamefont{Nesvizhevskii}},
  \bibinfo{author}{\bibfnamefont{A.}~\bibnamefont{Serebrov}},
  \bibinfo{author}{\bibfnamefont{A.}~\bibnamefont{Strelkov}},
  \bibinfo{author}{\bibfnamefont{S.}~\bibnamefont{Sumbaev}},
  \bibinfo{author}{\bibfnamefont{R.}~\bibnamefont{Tal’daev}},
  \bibnamefont{et~al.}, \bibinfo{journal}{JETP Lett}
  \textbf{\bibinfo{volume}{52}}, \bibinfo{pages}{373} (\bibinfo{year}{1990}).

\bibitem[{\citenamefont{Kossakowski et~al.}(1989)\citenamefont{Kossakowski,
  Grivot, Liaud, Schreckenbach, and Azuelos}}]{kossakowski89}
\bibinfo{author}{\bibfnamefont{R.}~\bibnamefont{Kossakowski}},
  \bibinfo{author}{\bibfnamefont{P.}~\bibnamefont{Grivot}},
  \bibinfo{author}{\bibfnamefont{P.}~\bibnamefont{Liaud}},
  \bibinfo{author}{\bibfnamefont{K.}~\bibnamefont{Schreckenbach}},
  \bibnamefont{and} \bibinfo{author}{\bibfnamefont{G.}~\bibnamefont{Azuelos}},
  \bibinfo{journal}{Nuclear Physics A} \textbf{\bibinfo{volume}{503}},
  \bibinfo{pages}{473} (\bibinfo{year}{1989}).

\bibitem[{\citenamefont{Paul et~al.}(1989)\citenamefont{Paul, Anton, Paul,
  Paul, and Mampe}}]{Paul89}
\bibinfo{author}{\bibfnamefont{W.}~\bibnamefont{Paul}},
  \bibinfo{author}{\bibfnamefont{F.}~\bibnamefont{Anton}},
  \bibinfo{author}{\bibfnamefont{L.}~\bibnamefont{Paul}},
  \bibinfo{author}{\bibfnamefont{S.}~\bibnamefont{Paul}}, \bibnamefont{and}
  \bibinfo{author}{\bibfnamefont{W.}~\bibnamefont{Mampe}},
  \bibinfo{journal}{Zeitschrift f{\"u}r Physik C Particles and Fields}
  \textbf{\bibinfo{volume}{45}}, \bibinfo{pages}{25} (\bibinfo{year}{1989}).

\bibitem[{\citenamefont{Last et~al.}(1988)\citenamefont{Last, Arnold,
  D{\"o}hner, Dubbers, and Freedman}}]{Last88}
\bibinfo{author}{\bibfnamefont{J.}~\bibnamefont{Last}},
  \bibinfo{author}{\bibfnamefont{M.}~\bibnamefont{Arnold}},
  \bibinfo{author}{\bibfnamefont{J.}~\bibnamefont{D{\"o}hner}},
  \bibinfo{author}{\bibfnamefont{D.}~\bibnamefont{Dubbers}}, \bibnamefont{and}
  \bibinfo{author}{\bibfnamefont{S.}~\bibnamefont{Freedman}},
  \bibinfo{journal}{Physical review letters} \textbf{\bibinfo{volume}{60}},
  \bibinfo{pages}{995} (\bibinfo{year}{1988}).

\bibitem[{\citenamefont{Spivak}(1988)}]{Spivak88}
\bibinfo{author}{\bibfnamefont{P.}~\bibnamefont{Spivak}},
  \bibinfo{journal}{Soviet Physics-JETP (English Translation)}
  \textbf{\bibinfo{volume}{67}}, \bibinfo{pages}{1735} (\bibinfo{year}{1988}).

\bibitem[{\citenamefont{Kosvintsev
  et~al.}(1986{\natexlab{a}})\citenamefont{Kosvintsev, Morozov, and
  Terekhov}}]{Kosvintsev86}
\bibinfo{author}{\bibfnamefont{Y.~Y.} \bibnamefont{Kosvintsev}},
  \bibinfo{author}{\bibfnamefont{V.}~\bibnamefont{Morozov}}, \bibnamefont{and}
  \bibinfo{author}{\bibfnamefont{G.}~\bibnamefont{Terekhov}},
  \bibinfo{journal}{JETP Letters} \textbf{\bibinfo{volume}{44}},
  \bibinfo{pages}{571} (\bibinfo{year}{1986}{\natexlab{a}}).

\bibitem[{\citenamefont{Kosvintsev
  et~al.}(1986{\natexlab{b}})\citenamefont{Kosvintsev, Morozov, and
  Terekhov}}]{Kosvintsev80}
\bibinfo{author}{\bibfnamefont{Y.~Y.} \bibnamefont{Kosvintsev}},
  \bibinfo{author}{\bibfnamefont{V.}~\bibnamefont{Morozov}}, \bibnamefont{and}
  \bibinfo{author}{\bibfnamefont{G.}~\bibnamefont{Terekhov}},
  \bibinfo{journal}{JETP Letters} \textbf{\bibinfo{volume}{44}},
  \bibinfo{pages}{571} (\bibinfo{year}{1986}{\natexlab{b}}).

\bibitem[{\citenamefont{Christensen et~al.}(1972)\citenamefont{Christensen,
  Nielsen, Bahnsen, Brown, and Rustad}}]{Christensen72}
\bibinfo{author}{\bibfnamefont{C.~J.} \bibnamefont{Christensen}},
  \bibinfo{author}{\bibfnamefont{A.}~\bibnamefont{Nielsen}},
  \bibinfo{author}{\bibfnamefont{A.}~\bibnamefont{Bahnsen}},
  \bibinfo{author}{\bibfnamefont{W.}~\bibnamefont{Brown}}, \bibnamefont{and}
  \bibinfo{author}{\bibfnamefont{B.}~\bibnamefont{Rustad}},
  \bibinfo{journal}{Physical Review D} \textbf{\bibinfo{volume}{5}},
  \bibinfo{pages}{1628} (\bibinfo{year}{1972}).

\bibitem[{\citenamefont{{Ivezi{\'c}} et~al.}(2014)\citenamefont{{Ivezi{\'c}},
  {Connolly}, {Vanderplas}, and {Gray}}}]{astroML}
\bibinfo{author}{\bibfnamefont{{\v Z}.}~\bibnamefont{{Ivezi{\'c}}}},
  \bibinfo{author}{\bibfnamefont{A.}~\bibnamefont{{Connolly}}},
  \bibinfo{author}{\bibfnamefont{J.}~\bibnamefont{{Vanderplas}}},
  \bibnamefont{and} \bibinfo{author}{\bibfnamefont{A.}~\bibnamefont{{Gray}}},
  \emph{\bibinfo{title}{Statistics, Data Mining and Machine Learning in
  Astronomy}} (\bibinfo{publisher}{Princeton University Press},
  \bibinfo{year}{2014}).

\bibitem[{\citenamefont{{Babu} and {Feigelson}}(2006)}]{Babu}
\bibinfo{author}{\bibfnamefont{G.~J.} \bibnamefont{{Babu}}} \bibnamefont{and}
  \bibinfo{author}{\bibfnamefont{E.~D.} \bibnamefont{{Feigelson}}}, in
  \emph{\bibinfo{booktitle}{Astronomical Data Analysis Software and Systems
  XV}}, edited by \bibinfo{editor}{\bibfnamefont{C.}~\bibnamefont{{Gabriel}}},
  \bibinfo{editor}{\bibfnamefont{C.}~\bibnamefont{{Arviset}}},
  \bibinfo{editor}{\bibfnamefont{D.}~\bibnamefont{{Ponz}}}, \bibnamefont{and}
  \bibinfo{editor}{\bibfnamefont{S.}~\bibnamefont{{Enrique}}}
  (\bibinfo{year}{2006}), vol. \bibinfo{volume}{351} of
  \emph{\bibinfo{series}{Astronomical Society of the Pacific Conference
  Series}}, p. \bibinfo{pages}{127}.

\end{thebibliography}
\end{document}